\documentclass{article}
\usepackage{spconf,hyperref}

\usepackage{comment}

\usepackage{cite}
\usepackage{amsmath,amssymb,amsfonts}
\usepackage{algorithmic}
\usepackage{graphicx}
\usepackage{textcomp}
\usepackage{xcolor}
\usepackage{multirow}
\usepackage{booktabs}
\usepackage{marvosym}
\usepackage{cite}
\usepackage{pifont}
\usepackage{setspace} 

\usepackage{makecell}
\usepackage[table]{xcolor}

\definecolor{lightgray}{gray}{0.9}

\title{FLOWSE-GRPO: TRAINING FLOW MATCHING SPEECH ENHANCEMENT VIA ONLINE REINFORCEMENT LEARNING}
\name{Haoxu Wang, Biao Tian, Yiheng Jiang, Zexu Pan, Shengkui Zhao, Bin Ma, Daren Chen, Xiangang Li}
\address{Tongyi Lab, Alibaba Group, China}
\begin{document}
\ninept
\maketitle
\begin{abstract}

Generative speech enhancement offers a promising alternative to traditional discriminative methods by modeling the distribution of clean speech conditioned on noisy inputs. Post‑training alignment via reinforcement learning (RL) effectively aligns generative models with human preferences and downstream metrics in domains such as natural language processing, but its use in speech enhancement remains limited, especially for online RL. Prior work explores offline methods like Direct Preference Optimization (DPO); online methods such as Group Relative Policy Optimization (GRPO) remain largely uninvestigated. In this paper, we present the first successful integration of online GRPO into a flow‑matching speech enhancement framework, enabling efficient post‑training alignment to perceptual and task‑oriented metrics with few update steps. Unlike prior GRPO work on Large Language Models, we adapt the algorithm to the continuous, time‑series nature of speech and to the dynamics of flow‑matching generative models. We show that optimizing a single reward yields rapid metric gains but often induces reward hacking that degrades audio fidelity despite higher scores. To mitigate this, we propose a multi‑metric reward optimization strategy that balances competing objectives, substantially reducing overfitting and improving overall performance. Our experiments validate online GRPO for speech enhancement and provide practical guidance for RL‑based post‑training of generative audio models.

\end{abstract}
\begin{keywords}
Group Relative Policy Optimization, Speech Enhancement, Post Training, Reinforcement Learning, Generative Models
\end{keywords}
\section{Introduction}
\label{sec:intro}
\vspace{-5pt}


Speech enhancement (SE) aims to estimate the original clean waveform from noisy audio, improving perceived quality and supporting downstream speech tasks. Past work mainly uses discriminative methods that predict the clean spectrum in the magnitude or complex domain to estimate an enhanced waveform \cite{mpsenet,zipenhancer}. Recently, generative SE has gained interest: borrowing from speech synthesis, these methods treat noisy audio as a conditioning input and model the distribution of clean speech, typically by maximizing the likelihood of the clean waveform \cite{wang2024selm,anyenhance,flowse}.

Current generative SE approaches can be grouped roughly into two main categories: discrete token based and flow-matching based methods. Discrete token based methods use autoregressive (AR) or non‑autoregressive (NAR) schemes that predict clean audio discrete tokens (e.g., SELM\cite{wang2024selm},  Genhancer\cite{genhancer}, GenSE\cite{gense}, LLaSE‑G1\cite{llaseg1}). Some masked generative models (MGM) predict masked tokens conditioned on unmasked ones, such as MaskSR\cite{MaskSR}, AnyEnhance\cite{anyenhance}. Flow-matching\cite{flowmatching} methods operate on mel‑spectrograms or latent representations and use flow-matching to predict the velocity field that transports a standard Gaussian to the clean audio distribution\cite{flowavse,flowse}.


Meanwhile, post‑training effectively aligns models to human perception and downstream objectives and is widely applied in generative domains, including language understanding\cite{ouyang2022training}, image generation\cite{xu2023imagereward}, and speech synthesis\cite{gao25d_interspeech,sun2025f5r}. It finetunes pretrained generative backbones using reinforcement learning (RL)‑style or preference‑alignment objectives to improve human‑centric metrics. For example, \cite{gao25d_interspeech} applies differentiable optimization to CosyVoice2\cite{du2024cosyvoice2}’s Large Language Models (LLM) to improve intelligibility and emotional expressiveness; F5R‑TTS\cite{sun2025f5r} uses Group Relative Policy Optimization (GRPO)\cite{shao2024deepseekmath} to reduce word error rate and improve speaker similarity. \cite{sedpo} uses the offline RL method Direct Preference Optimization (DPO)\cite{dpo} to optimize the restoration metrics in speech restoration.


\begin{figure*}[!t]
	\centering
\includegraphics[width=0.80\linewidth]{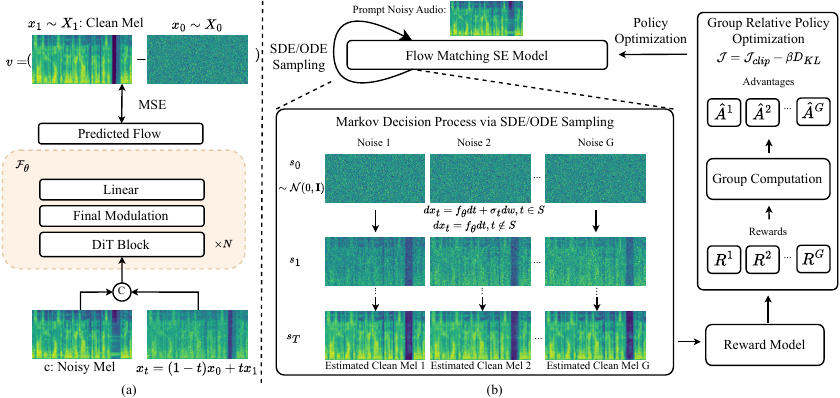}
        \vspace{-10pt}
	\caption{
        (a) The structure of our Flow matching based speech enhancement model. (b) The pipeline of post-training using GRPO.
	}
	\label{fig:grpo}
        \vspace{-20pt}
\end{figure*}


However, \cite{sedpo} still uses DPO, an off‑policy method. GRPO is an on‑policy algorithm that more effectively finetunes models and can optimize diverse metrics without offline generation of large numbers of win–lose pairs. Flow‑GRPO \cite{flowgrpo} first introduces online RL to flow‑matching models by converting the original Ordinary Differential Equation (ODE) trajectory into a Stochastic Differential Equation (SDE) to provide the stochasticity required by on‑policy RL, and demonstrates improved human‑perceptual metrics in image generation.
In this paper, we present the first successful integration of online GRPO into a flow‑matching SE framework, as an on‑policy RL post‑training procedure without modifying their original architectures (unlike F5R‑TTS\cite{sun2025f5r}).
This online RL method can finetune the base model and improve perceptual quality and objective metrics such as DNSMOS\cite{reddy2022dnsmos835}, SpeechBERTScore\cite{speechbertscore}, and speaker similarity. We systematically study GRPO training configurations for enhancement metrics and find that optimizing a single reward rapidly improves that metric but can induce reward hacking that degrades others; a multi‑metric reward mitigates this trade‑off. Overall, we validate on‑policy GRPO for speech enhancement and provide practical guidance for future post‑training of generative enhancement models.


\section{Methods}
\vspace{-5pt}

\subsection{Flow-matching based speech enhancement}
\vspace{-5pt}

We train a flow‑matching SE system with a pretrained vocoder. The flow‑matching model predicts the velocity field that transports a base gaussian distribution to the clean mel‑spectrogram, conditioned on the noisy mel‑spectrogram. The pretrained vocoder then converts the estimated clean mel‑spectrogram back to the waveform.

Specifically, as shown in Fig. \ref{fig:grpo}(a), the flow‑matching model aims to predict $v = x_{1} - x_{0}$ from the input $x_{t} = (1 - t)x_{0} + t x_{1}$, conditioned on the noisy mel $c$. Here $x_{1} \sim X_{1}$  is sampled from the clean mel‑spectrogram distribution and $x_{0} \sim X_{0} = \mathcal{N}(0,\mathbf{I})$. The model's optimization objective $\mathcal{L}(\theta)$ is defined as:

\begin{equation}
\begin{array}{ccc}
\mathcal{L}(\theta) = E_{t, x_{0} \sim X_{0}, x_{1} \sim X_{1}} [|| v - v_{\theta}(x_{t}, c, t)|| ^{2}]
\end{array}
\end{equation}

We concatenate $x_{t}$ and noisy mel $c$ along the channel dimension and train $\theta$ with loss $\mathcal{L}(\theta)$. The flow‑matching backbone is DiT\cite{dit} and the vocoder is the pretrained HiFi‑GAN\cite{hifigan} from \cite{du2024cosyvoice2}.


Note that in the original Flow‑GRPO \cite{flowgrpo}, the time of the denoising steps is $t: 1 \to 0$. In the flow-matching-based SE task, the denoising time proceeds in the opposite direction: $t: 0 \to 1$.
To reconcile these conventions, we set new $v = -v_{se}$, $t = 1 - t_{se}$, $x_{0} = x_{se, t=1} \sim X_{1}$, $x_{1} = x_{se, t=0} \sim X_{0}$, and $\Delta t = -\Delta t_{se}$.

\subsection{Flow-GRPO}
\vspace{-5pt}

\subsubsection{Markov Decision Process}
\vspace{-5pt}


Following Flow‑GRPO, the flow-matching decoding process can be formulated as a Markov decision process $(S,A,\rho_{0},P,R)$. Time runs from $1 \to 0$ over $1 \to T$ steps. At step $t$ the state is $s_{t}\triangleq (c,t,x_{t})$ and the next state is $s_{t+1}\triangleq (c,t,x_{t-1})$. The action $a_{t}\triangleq x_{t-1}$ is the denoised sample produced from the model‑predicted velocity, and the policy is $\pi(a_{t}\mid s_{t})=p_{\theta}(x_{t-1}\mid x_{t},c)$. Transitions are deterministic: $P(s_{t+1}\mid s_{t},a_{t})\triangleq (\delta_{c},\delta_{t-1},\delta_{x_{t-1}})$, where $\delta_{y}$ is a Dirac delta centered at $y$. The initial state distribution is $\rho_{0}(s_{0})\triangleq (p_{c},\delta_{T},\mathcal{N}(0,\mathbf{I}))$. A reward is given only at $t=0$ (after getting the final mel and decoding it to waveform): $R(s_{t},a_{t})\triangleq r(x_{0},c)$ if $t=0$, and $0$ otherwise.

\subsubsection{ODE to SDE}
\vspace{-5pt}

The standard flow‑matching denoising process is a deterministic ODE sampler and therefore cannot produce diverse samples by stochastic sampling, making it unsuitable for on‑policy RL. Flow‑GRPO converts this deterministic ODE sampler into an equivalent SDE sampler that preserves the same marginal distributions while introducing the stochasticity required by GRPO. The standard flow‑matching model is solved using the deterministic ODE:

\vspace{-5pt}
\begin{equation}
\begin{array}{ccc}
dx_{t} = v_{t}dt
\end{array}
\end{equation}
\vspace{-5pt}


Flow‑GRPO converts the deterministic ODE into an equivalent SDE that preserves the model’s marginal probability densities at all time steps. The resulting reverse‑time SDE is:

\vspace{-10pt}
\begin{equation}
\begin{array}{ccc}
dx_{t} = (v_{t}(x_{t}) - \frac{\sigma_{t}^{2}}{2} \nabla logp_{t}(x_{t}))dt + \sigma_{t} dw
\end{array}
\end{equation}
\vspace{-10pt}

According to \cite{flowgrpo}, the above expression can be transformed into:

\vspace{-5pt}
\begin{equation}
\begin{array}{ccc}
dx_{t} = [v_{t}(x_{t}) + \frac{\sigma_{t}^{2}}{2(1-t)}(x_{t} + tv_{t}(x_{t}))dt ] + \sigma_{t} dw
\end{array}
\end{equation}
\vspace{-5pt}

The final update rule in \cite{flowgrpo} is:

\vspace{-10pt}
\begin{equation}
\begin{array}{ccc}
x_{t + \Delta t} = x_{t, \mathrm{mean}} + \sigma_{t} \sqrt{\Delta t} \epsilon \\
x_{t, \mathrm{mean}} = x_{t} + [ v_{\theta}(x_{t}, t) + \frac{\sigma_{t}^{2}}{2t} (x_{t} + (1 - t)v_{\theta}(x_{t}, t)) ] \Delta t
\end{array}
\end{equation}
\vspace{-10pt}


Because our flow‑matching SE uses the opposite time ordering to the original Flow-GRPO, we substitute symbols accordingly. All $t$ in equation \ref{eq:sde} correspond to $t_{se}\!:\,0\to1$. In the following, $t$ denotes $t_{se}$ and we omit the subscript “se” for brevity.

\vspace{-10pt}
\begin{equation}
\label{eq:sde}
\begin{array}{rcl}
x_{t, \mathrm{mean}} &=& x_{t} + \big[ -v_{\theta}(x_{t}, t) + \frac{\sigma_{t}^{2}}{2(1-t)} (x_{t} - t v_{\theta}(x_{t}, t)) \big] (-\Delta t) \\
&=& x_{t} + \big[ v_{\theta}(x_{t}, t) + \frac{\sigma_{t}^{2}}{2(1-t)} (-x_{t} + t v_{\theta}(x_{t}, t)) \big] \Delta t
\end{array}
\end{equation}
\vspace{-10pt}

\noindent where $\epsilon\sim\mathcal{N}(0,\mathbf{I})$ and $\sigma_{t}=a\sqrt{\frac{1-t}{t}}$. The noise level $a$ is a hyperparameter that controls stochasticity in the reverse‑time denoising process.


We also follow MixGRPO\cite{li2025mixgrpo} and Flow‑GRPO‑Fast\cite{flowgrpo} and use window training: we apply SDE sampling only to a subset of early steps $t \in S = [S_{min}, S_{min} + ws]$, where $S_{min}$ denotes the SDE window start and $ws$ the window size, while remaining steps use deterministic ODE updates. Optimization is performed only on these stochastic steps to reduce training burden and accelerate convergence.

\vspace{-10pt}
\subsubsection{GRPO Loss}
\vspace{-5pt}
RL learns a policy that maximizes the expected cumulative reward. The policy optimization objective is defined as:

\vspace{-10pt}
\begin{equation}
\begin{array}{ccc}
\mathrm{max}_{\theta} E_{(s_{0}, a_{0}, ...) \sim \pi_{\theta}} \big[ \sum_{t \in S} (R(s_{t}, a_{t}) \\ - \beta D_{KL}(\pi_{\theta}(\cdot | s_{t}) || \pi_{ref}(\cdot | s_{t})  ) )\big]
\end{array}
\end{equation}
\vspace{-10pt}


GRPO estimates the advantage $A$ using an intra‑group relative formulation. As shown in Fig \ref{fig:grpo}(b), given a noisy prompt $c$, the flow‑matching SE model $p_{\theta}$ samples a group of $G$ candidate clean outputs $\{\hat{x}_{T=1}^{i}\}_{i=1}^{G}$ with the mixed ODE–SDE sampler, together with their trajectories $\{(x_{0}^{i},\ldots,x_{T-1}^{i},x_{T=1}^{i})\}_{i=1}^{G}$. The advantage for the $i$‑th estimated clean audio is then computed from the group‑wise policy rewards.

\begin{equation}
\begin{array}{ccc}
\hat A_{t}^{i} = \frac{ R(\hat x_{1}^{i}, c) - \mathrm{mean}(\{R(\hat x_{1}^{i}, c)\}_{i=1}^{G}) } { \mathrm{std}(\{R(\hat x_{1}^{i}, c)\}_{i=1}^{G}) }
\end{array}
\end{equation}

GRPO updates the policy model by maximizing the following objective:

\begin{equation}
\begin{array}{ccc}
\mathcal{J}_{\mathrm{flow\text{-}grpo}}(\theta) = E_{c \sim C, \{x_{i}\}_{i=1}^{G} \sim \pi_{\theta_{\mathrm{old}}(\cdot | c) } }f(r, \hat A, \theta, \epsilon, \beta), \\

f(r, \hat A, \theta, \epsilon, \beta) = \frac{1}{G} \sum_{i=1}^{G} \frac{1}{S} \sum_{t \in S} \big(  \\ 
\mathrm{min}( r_{t}^{i}(\theta)\hat A_{t}^{i}, \mathrm{clip}(r_{t}^{i}(\theta), 1-\epsilon, 1+\epsilon) \hat A_{t}^{i} ) \big) \\
- \beta D_{\mathrm{KL}}(\pi_{\theta} || \pi_{\theta_{\mathrm{ref}}}) )

\end{array}
\end{equation}

\noindent where
$
r_{t}^{i}(\theta)=\frac{p_{\theta}(x_{t+1}^{i}\mid x_{t}^{i},c)}{p_{\theta_{\mathrm{old}}}(x_{t+1}^{i}\mid x_{t}^{i},c)}.
$
The likelihood $p_{\theta}(x_{t+1}^{i}\mid x_{t}^{i},c)$ is the gaussian density of $x_{t+\Delta t}$ with mean $x_{t,\mathrm{mean}}$ and standard deviation $\sigma_{t}\sqrt{\Delta t}$. Maximizing $\mathcal{J}_{\mathrm{flow\text{-}grpo}}(\theta)$ updates the model by optimizing $p_{\theta}(x_{t+1}^{i}\mid x_{t}^{i},c)$.



\begin{figure*}[!t]
	\centering
\includegraphics[width=0.80\linewidth]{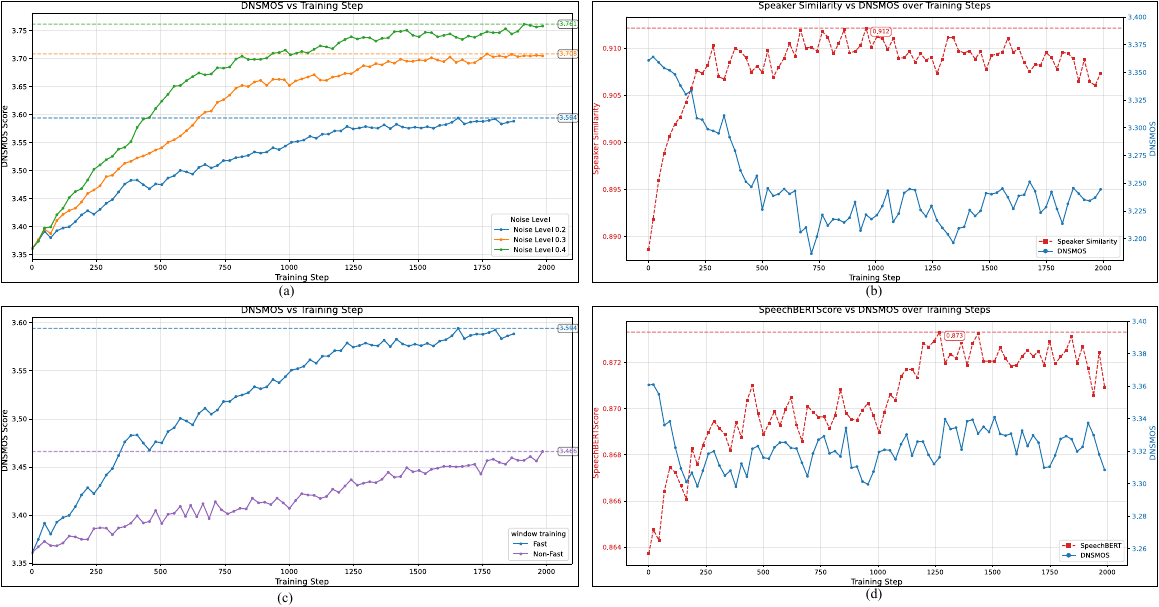}
    \vspace{-10pt}
	\caption{
        (a) The DNSMOS vs training steps with different Noise Level. (b) The Speaker Similarity vs training steps. (c) Effect of window training.  (d) The SpeechBERTScore vs training steps. All results are evaluated on the DNS2020 No Reverb test set.
	}
	\label{fig:figs}
    \vspace{-15pt}
\end{figure*}

\subsection{Reward of the SE task}

\textbf{DNSMOS}: A non‑intrusive metric that estimates audio quality. It predicts scores based on ITU‑T P.835 and outputs three measures: speech quality (SIG), background noise quality (BAK), and overall quality (OVRL). This metric commonly evaluates discriminative and generative SE when no reference is available. We use $\frac{1}{4}$ weighted OVRL as the reward $\frac{R_{\mathrm{DNSMOS}}}{4}$ for GRPO training. All audio is resampled to 16 kHz before evaluation.


\textbf{Speaker similarity} measures the speaker correlation between the estimated clean audio and the reference. We extract speaker embeddings from both signals using the same ERes2Net\cite{ERes2Net} model as CosyVoice2\cite{du2024cosyvoice2} and compute their cosine similarity as the speaker similarity score. The score ranges in $[0,1]$ (higher is better) and serves as the reward $R_{\mathrm{Spk}}$ in GRPO training.

\begin{table}[t]\centering
\small
\setlength{\tabcolsep}{0.8mm}{
  \centering
  \footnotesize
  \caption{\label{tab:previous} {Evaluation results on the DNS2020 challenge test set. 
}}
  \begin{tabular}{ccccccc}
  \toprule
  \textbf{Data} & Model & Type & RL & SIG & BAK & OVRL \\
  
  \midrule

  \multirow{13}*{No Reverb} & MaskSR & MGM & \ding{55} & 3.586 & 4.116 & 3.339 \\
  & GenSE & AR & \ding{55} & 3.650 & 4.180 & 3.430 \\
  & LLaSE-G1 & AR & \ding{55} & 3.660 & 4.170 & 3.420 \\
  & FlowSE & FM & \ding{55} & 3.685 & 4.201 & 3.445 \\
  \cmidrule(lr){2-7}
  & \multirow{2}*{AnyEnhance} & \multirow{2}*{MGM} & \ding{55} & 3.640 & 4.179 & 3.418 \\
  &  &  & DPO & 3.684 & 4.203 & 3.476 \\
  & \multirow{2}*{AR+Soundstorm} & \multirow{2}*{AR} & \ding{55} & 3.648 & 4.155 & 3.422 \\
  &  &  & DPO & 3.673 & 4.194 & 3.469 \\
  & \multirow{3}*{Flow-SR} & \multirow{3}*{FM} & \ding{55} & 3.581 & 4.133 & 3.355 \\
  &  &  & \multirow{2}*{DPO} & 3.632 & 4.173 & 3.420 \\
&  &  &  & 
  \vspace{-0.4ex}\raisebox{+0.5ex}{{\scriptsize\textcolor{red}{\hspace{0.4em}(+0.051)}}} &
  \vspace{-0.4ex}\raisebox{+0.5ex}{{\scriptsize\textcolor{red}{\hspace{0.4em}(+0.040)}}} &
  \vspace{-0.4ex}\raisebox{+0.5ex}{{\scriptsize\textcolor{red}{\hspace{0.4em}(+0.065)}}} \\

  \cmidrule(lr){2-7}
  
  & \multirow{3}*{\parbox{1.8cm}{\centering FlowSE-GRPO \\ (Ours)}} & \multirow{3}*{FM} & \ding{55} & 3.598 &	4.172 & 3.373 \\

  &  &  & \multirow{2}*{GRPO} & \textbf{3.753} & \textbf{4.248} & \textbf{3.549} \\
&  &  &  & 
  \vspace{-0.4ex}\raisebox{+0.5ex}{{\scriptsize\textcolor{red}{\hspace{0.4em}(+0.155)}}} &
  \vspace{-0.4ex}\raisebox{+0.5ex}{{\scriptsize\textcolor{red}{\hspace{0.4em}(+0.076)}}} &
  \vspace{-0.4ex}\raisebox{+0.5ex}{{\scriptsize\textcolor{red}{\hspace{0.4em}(+0.176)}}} \\
  \midrule

  \multirow{13}*{With Reverb} & MaskSR & MGM & \ding{55} & 3.531 & 4.065 & 3.253 \\
  & GenSE & AR & \ding{55} & 3.490 & 3.730 & 3.190 \\
  & LLaSE-G1 & AR & \ding{55} & 3.590 & 4.100 & 3.330 \\
  & FlowSE & FM & \ding{55} & 3.601 & 4.102 & 3.331 \\
  \cmidrule(lr){2-7}
  & \multirow{2}*{AnyEnhance} & \multirow{2}*{MGM} & \ding{55} & 3.500 & 4.040 & 3.204 \\
  &  &  & DPO & 3.670 & 4.178 & 3.438 \\
  & \multirow{2}*{AR+Soundstorm} & \multirow{2}*{AR} & \ding{55} & 3.681 & 4.127 & 3.431 \\
  &  &  & DPO &  3.709 & 4.189 & 3.496 \\
  & \multirow{3}*{Flow-SR} & \multirow{3}*{FM} & \ding{55} & 3.539 & 4.019 & 3.255 \\
  &  &  & \multirow{2}*{DPO} & 3.629 & 4.163 & 3.399 \\
&  &  &  & 
  \vspace{-0.4ex}\raisebox{+0.5ex}{{\scriptsize\textcolor{red}{\hspace{0.4em}(+0.090)}}} &
  \vspace{-0.4ex}\raisebox{+0.5ex}{{\scriptsize\textcolor{red}{\hspace{0.4em}(+0.144)}}} &
  \vspace{-0.4ex}\raisebox{+0.5ex}{{\scriptsize\textcolor{red}{\hspace{0.4em}(+0.144)}}} \\

  \cmidrule(lr){2-7}
  & \multirow{3}*{\parbox{1.8cm}{\centering FlowSE-GRPO \\ (Ours)}} & \multirow{3}*{FM} & \ding{55} & 3.511 &	4.105 & 3.254 \\


&  &  & \multirow{2}*{GRPO} & \textbf{3.740} & \textbf{4.251} & \textbf{3.530} \\
&  &  &  & 
  \vspace{-0.4ex}\raisebox{+0.5ex}{{\scriptsize\textcolor{red}{\hspace{0.4em}(+0.229)}}} &
  \vspace{-0.4ex}\raisebox{+0.5ex}{{\scriptsize\textcolor{red}{\hspace{0.4em}(+0.146)}}} &
  \vspace{-0.4ex}\raisebox{+0.5ex}{{\scriptsize\textcolor{red}{\hspace{0.4em}(+0.276)}}} \\




  \bottomrule

\end{tabular}
}
\vspace{-20pt}
\end{table}

\textbf{SpeechBERTScore}: A reference‑based metric for semantic similarity between two speech signals. It computes BERTScore on self‑supervised dense speech features extracted from the generated and reference utterances; in practice, it evaluates precision over discretized speech tokens. We use SpeechBERTScore to measure semantic similarity between the estimated clean audio and the reference, and as the reward $R_{\mathrm{Speechbertscore}}$ in GRPO training. The score ranges in $[0,1]$ (higher is better).


\textbf{Multi‑metric reward}: To balance each metric’s influence, we normalize them by their standard deviation computed from collected data and combine them with a weighted sum as the final reward:
$
R = \lambda_{1} \frac{R_{\mathrm{DNSMOS}}}{\mathrm{std}(R_{\mathrm{DNSMOS}})} + \lambda_{2} \frac{R_{\mathrm{Spk}}}{\mathrm{std}(R_{\mathrm{Spk}})} + \lambda_{3} \frac{R_{\mathrm{Speechbertscore}}}{\mathrm{std}(R_{\mathrm{Speechbertscore}})}.
$

\begin{table}[t]\centering
\small
\setlength{\tabcolsep}{0.8mm}{
  \centering
  \footnotesize
  \caption{\label{tab:compare} {Comparisons with other previous methods on DNS 2020 challenge. SPK and SBS represent Speaker similarity and SpeechBERTScore. 
}}
  \begin{tabular}{ccccccc}
  \toprule
  \textbf{Data}  & RL & SIG & BAK & OVRL & SPK[\%] & SBS[\%] \\
  
  \midrule


  \multirow{2}*{No Reverb}   & \ding{55} & 3.598 & 4.172 & 3.373 & 88.88 & 86.35 \\

  &  \cellcolor{lightgray}GRPO & \cellcolor{lightgray}3.753 & \cellcolor{lightgray}4.248  & \cellcolor{lightgray}3.549 & \cellcolor{lightgray}90.43 & \cellcolor{lightgray}86.72 \\

  \midrule

  \multirow{2}*{With Reverb}   & \ding{55} & 3.511 & 4.105 & 3.254 & 73.72 & 73.62 \\

  &  \cellcolor{lightgray}GRPO & \cellcolor{lightgray}3.740 & \cellcolor{lightgray}4.251 & \cellcolor{lightgray}3.530 & \cellcolor{lightgray}77.75 & \cellcolor{lightgray}75.89 \\

  \midrule

  \multirow{2}*{Real Recording}   & \ding{55} & 3.397 &	4.035 & 3.115 & - & - \\

  &  \cellcolor{lightgray}GRPO & \cellcolor{lightgray}3.604	& \cellcolor{lightgray}4.161 & \cellcolor{lightgray}3.356 & \cellcolor{lightgray}- & \cellcolor{lightgray}- \\

  \bottomrule

\end{tabular}
}
\vspace{-20pt}
\end{table}





\section{Experimental Setup}
\vspace{-10pt}

\subsection{Dataset}
\vspace{-5pt}


Our training has two stages: Pre‑Training and Post‑Training. We apply GRPO during the Post‑Training stage. 
For Pre‑Training, clean speech consists of DNS2020 (clean English)\cite{dns2020}, LibriTTS‑960\cite{libritts}, VCTK\cite{voicebank}, WSJ\cite{deepcluster}, EARS\cite{ears}, and mls\_english (Track 2 of the Urgent Challenge 2025)\cite{urgent2025}. Noise sources include the DNS2020 noise set\cite{dns2020}, WHAM!\cite{wham}, DEMAND\cite{demands}, fma\_medium\cite{fma}, FSD50k (excluding speech labels)\cite{fsd50k}, simulated wind noise\cite{urgent2025}, and noises from RealMan\cite{realman}. Reverberation uses OpenSLR26 and OpenSLR28\cite{openslr2628}. For Post‑Training, we use only LibriTTS‑960 as clean speech; all noise sources above form the training set. For model evaluation, we use the non-blind test sets (No Reverb, With Reverb, Real Recording) provided by DNS2020 Challenge. 

\vspace{-5pt}
\subsection{Training and Model Configuration}
\vspace{-5pt}


The FlowSE‑GRPO backbone uses DiT\cite{dit} with hidden dimension 512, 12 layers, 8 attention heads, and FFN hidden size 1024, totaling 46.18M parameters. In the Pre‑Training stage, the initial learning rate (LR) is 1e‑4 with a 10k‑step warmup followed by linear decay. Training uses 4 GPUs with dynamic batching, each batch containing 100s of audio per GPU. The model is trained for 100k steps. In Post‑Training, we apply Low-Rank Adaptation (LoRA) with rank 32 and $\alpha = 64$, yielding 1.57M trainable parameters. The LR is 2e‑4 with no warmup and a linear decay schedule.




During each Post‑Training iteration, we perform data collection followed by on‑policy training. For data collection, we sample 6 noisy prompt wavs and repeat this 12 times to obtain 72 prompts. For each prompt, we generate $G = 10$ SDE different samples, producing 720 candidates, which we score. We discard any group with zero within‑group standard deviation. For on‑policy training, the remaining samples are regrouped into batches of size $\mathrm{batchsize}=12$. Each iteration performs 4 parameter updates, each update counted as one training step.

We use classifier‑free guidance during both training and inference. During training, we apply SDE sampling to only two time steps $ws = 2$; the denoising step is sampled from [7,10] and $S_{min}$ from [1,3]. In inference, the denoising step is set to 10.


\section{Results and Discussion}
\vspace{-5pt}

\subsection{Main Results}
\vspace{-5pt}

\subsubsection{Results of single‑metric reward optimization}


Fig. \ref{fig:figs}(a) shows the effect of optimizing only the DNSMOS metric. Focusing on the blue curve ($a=0.2$), DNSMOS rises from about 3.36 and, after roughly 1.5k steps, converges near 3.59. This demonstrates GRPO's ability to align the base model with downstream perceptual metrics. However, we observe reward hacking accompanying this gain; see Sec. \ref{sec:multiresults} for discussion.


Fig. \ref{fig:figs}(b) shows the effect of using only the $R_{Spk}$ as the reward. Focusing on the red curve, speaker similarity increases from 88.86 to just above 91.20 during training (similarly to DNSMOS). Fig. \ref{fig:figs}(d) shows the effect of using only SpeechBERTScore as the reward. Focusing on the red curve, SpeechBERTScore rises from about 86.4 to about 87.2 during training. These demonstrate that online RL can optimize downstream metrics.



\subsubsection{Results of multi‑metric reward optimization}
\label{sec:multiresults}


As shown in Fig \ref{fig:figs}(a), (b), (d), we observe an implausible increase of DNSMOS to 3.59, and also find that optimizing only SpeechBERTScore or speaker similarity with online RL reduces DNSMOS, indicating reward hacking toward a single metric. To mitigate this, we combine DNSMOS, SpeechBERTScore, and speaker similarity into a multi‑metric reward with weights $\lambda_{1} = 0.6, \lambda_{2} = \lambda_{3} = 1$ (In early experiments, we try DNSMOS weights 1.0/0.6/0.5/0.4: too large over-focuses DNSMOS, too small under-optimizes. We use $\lambda_{1} = 0.6$ as a stable trade-off), set $a = 0.4$, and train for 5k steps.


Final results are reported in Table \ref{tab:compare}. On the DNS2020 No Reverb test set, compared to the base model without RL, the multi‑metric optimized model improves OVRL from 3.373 to 3.549, speaker similarity from 88.88 to 90.43, and SpeechBERTScore from 86.35 to 86.72. Additionally, on the DNS2020 With Reverb test set OVRL increases from 3.254 to 3.530, and on the real recording test set OVRL increases from 3.115 to 3.356. These results indicate that multi‑metric optimization alleviates reward hacking. We also find that score weights require careful tuning; we will investigate more robust online RL training strategies in future work.


As shown in Table \ref{tab:previous}, our base model without RL achieves performance comparable to Flow‑SR. Compared with DPO (+0.065 on No Reverb set, +0.144 on With Reverb set), GRPO more effectively improves downstream DNSMOS (+0.176 on No Reverb set, +0.274 on With Reverb set) via an on‑policy RL mechanism using fewer training steps (GRPO: 5k steps, DPO: 20k steps), demonstrating the superiority of online RL.


\subsection{Ablation Study}

\subsubsection{Effect of Noise Level}


Increasing the noise\_level $a$ expands the SDE exploration, which aids RL exploration. As shown in Fig. \ref{fig:figs}(a), raising $a$ from 0.2 to 0.3 and 0.4 increases the DNSMOS improvement rate during training, indicating faster learning from wider exploration. However, a larger $a$ may cause reward hacking, so the exploration range must be carefully balanced.

\subsubsection{Effect of window training}


As shown in Fig. \ref{fig:figs}(c), we perform an ablation study of window training. Training across all denoising steps ($S_{min}=1, ws=10$) (Non-Fast) yields slower reward growth than early‑step window training (Fast), indicating that applying RL only to early denoising steps accelerates reward improvement and speeds up model training.

\vspace{-5pt}
\section{Conclusion}
\vspace{-5pt}

We introduce GRPO into a generative speech enhancement framework. Unlike prior GRPO work on LLMs or offline DPO applied to flow‑matching, we integrate online RL into a flow‑matching SE model, enabling post‑training alignment to human perceptual and downstream task metrics with few training steps. We evaluate FlowSE‑GRPO across training configurations and policies and show consistent improvements on DNSMOS, speaker similarity, and SpeechBERTScore. We also find that optimizing a single metric can induce reward hacking and mitigate this by optimizing a composite multi‑metric reward. Overall, our results validate online policy GRPO for SE and offer practical guidance for post‑training generative SE models.


\begin{spacing}{0.3}
\footnotesize
\bibliographystyle{IEEEbib}
\bibliography{strings,refs,mybib}

@inproceedings{wang2024selm,
  title={{SELM: Speech Enhancement Using Discrete Tokens and Language Models}},
  author={Wang, Ziqian and Zhu, Xinfa and Zhang, Zihan and Lv, YuanJun and Jiang, Ning and Zhao, Guoqing and Xie, Lei},
  booktitle={Proc. ICASSP},
  pages={11561--11565},
  year={2024},
  organization={IEEE}
}

@inproceedings{MaskSR,
  title     = {{MaskSR: Masked Language Model for Full-band Speech Restoration}},
  author    = {Xu Li and Qirui Wang and Xiaoyu Liu},
  year      = {2024},
  booktitle = {Proc. Interspeech},
  pages     = {2275--2279},
  doi       = {10.21437/Interspeech.2024-1584},
  issn      = {2958-1796},
}

@inproceedings{genhancer,
  title     = {{Genhancer: High-Fidelity Speech Enhancement via Generative Modeling on Discrete Codec Tokens}},
  author    = {Haici Yang and Jiaqi Su and Minje Kim and Zeyu Jin},
  year      = {2024},
  booktitle = {Proc. Interspeech},
  pages     = {1170--1174},
  doi       = {10.21437/Interspeech.2024-590},
  issn      = {2958-1796},
}

@inproceedings{gense,
  author       = {Jixun Yao and
                  Hexin Liu and
                  Chen Chen and
                  Yuchen Hu and
                  Eng Siong Chng and
                  Lei Xie},
  title        = {GenSE: Generative Speech Enhancement via Language Models using Hierarchical
                  Modeling},
  booktitle    = {Proc. ICLR},
  publisher    = {OpenReview.net},
  year         = {2025},
  url          = {https://openreview.net/forum?id=1p6xFLBU4J},
  bibsource    = {dblp computer science bibliography, https://dblp.org}
}

@inproceedings{llaseg1,
  author       = {Boyi Kang and
                  Xinfa Zhu and
                  Zihan Zhang and
                  Zhen Ye and
                  Mingshuai Liu and
                  Ziqian Wang and
                  Yike Zhu and
                  Guobin Ma and
                  Jun Chen and
                  Longshuai Xiao and
                  Chao Weng and
                  Wei Xue and
                  Lei Xie},
  title        = {LLaSE-G1: Incentivizing Generalization Capability for LLaMA-based
                  Speech Enhancement},
  booktitle    = {Proc. ACL},
  pages        = {13292--13305},
  year         = {2025},
  url          = {https://aclanthology.org/2025.acl-long.651/},
  bibsource    = {dblp computer science bibliography, https://dblp.org}
}

@ARTICLE{anyenhance,
  author={Zhang, Junan and Yang, Jing and Fang, Zihao and Wang, Yuancheng and Zhang, Zehua and Wang, Zhuo and Fan, Fan and Wu, Zhizheng},
  journal={IEEE Transactions on Audio, Speech and Language Processing}, 
  title={AnyEnhance: A Unified Generative Model With Prompt-Guidance and Self-Critic for Voice Enhancement}, 
  year={2025},
  volume={33},
  number={},
  pages={3085-3098},
  doi={10.1109/TASLPRO.2025.3587393}}

@inproceedings{flowmatching,
  author       = {Xingchao Liu and
                  Chengyue Gong and
                  Qiang Liu},
  title        = {Flow Straight and Fast: Learning to Generate and Transfer Data with
                  Rectified Flow},
  booktitle    = {Proc. ICLR},
  publisher    = {OpenReview.net},
  year         = {2023},
  url          = {https://openreview.net/forum?id=XVjTT1nw5z},
  bibsource    = {dblp computer science bibliography, https://dblp.org}
}

@inproceedings{flowse,
  title     = {{FlowSE: Efficient and High-Quality Speech Enhancement via Flow Matching}},
  author    = {Ziqian Wang and Zikai Liu and Xinfa Zhu and Yike Zhu and Mingshuai Liu and Jun Chen and Longshuai Xiao and Chao Weng and Lei Xie},
  year      = {2025},
  booktitle = {Proc. Interspeech},
  pages     = {4858--4862},
  doi       = {10.21437/Interspeech.2025-1745},
  issn      = {2958-1796},
}

@inproceedings{flowavse,
  title     = {{FlowAVSE: Efficient Audio-Visual Speech Enhancement with Conditional Flow Matching}},
  author    = {Chaeyoung Jung and Suyeon Lee and Ji-Hoon Kim and Joon Son Chung},
  year      = {2024},
  booktitle = {Proc. Interspeech},
  pages     = {2210--2214},
  doi       = {10.21437/Interspeech.2024-701},
  issn      = {2958-1796},
}

@article{ouyang2022training,
  title={Training language models to follow instructions with human feedback},
  author={Ouyang, Long and Wu, Jeffrey and Jiang, Xu and Almeida, Diogo and Wainwright, Carroll and Mishkin, Pamela and Zhang, Chong and Agarwal, Sandhini and Slama, Katarina and Ray, Alex and others},
  journal={Proc. NeurIPS},
  volume={35},
  pages={27730--27744},
  year={2022}
}

@article{xu2023imagereward,
  title={Imagereward: Learning and evaluating human preferences for text-to-image generation},
  author={Xu, Jiazheng and Liu, Xiao and Wu, Yuchen and Tong, Yuxuan and Li, Qinkai and Ding, Ming and Tang, Jie and Dong, Yuxiao},
  journal={Proc. NeurIPS},
  volume={36},
  pages={15903--15935},
  year={2023}
}

@inproceedings{gao25d_interspeech,
  title     = {{Differentiable Reward Optimization for LLM based TTS system}},
  author    = {Changfeng Gao and Zhihao Du and Shiliang Zhang},
  year      = {2025},
  booktitle = {Proc. Interspeech},
  pages     = {2450--2454},
  doi       = {10.21437/Interspeech.2025-704},
  issn      = {2958-1796},
}

@article{sun2025f5r,
  title={F5R-TTS: Improving flow-matching based text-to-speech with group relative policy optimization},
  author={Sun, Xiaohui and Xiao, Ruitong and Mo, Jianye and Wu, Bowen and Yu, Qun and Wang, Baoxun},
  journal={arXiv preprint arXiv:2504.02407},
  year={2025}
}

@article{sedpo,
  title={Multi-Metric Preference Alignment for Generative Speech Restoration},
  author={Zhang, Junan and Zhang, Xueyao and Yang, Jing and Wang, Yuancheng and Fan, Fan and Wu, Zhizheng},
  journal={arXiv preprint arXiv:2508.17229},
  year={2025}
}

@article{dpo,
  title={Direct preference optimization: Your language model is secretly a reward model},
  author={Rafailov, Rafael and Sharma, Archit and Mitchell, Eric and Manning, Christopher D and Ermon, Stefano and Finn, Chelsea},
  journal={Proc. NeurIPS},
  volume={36},
  pages={53728--53741},
  year={2023}
}

@article{du2024cosyvoice2,
  title={Cosyvoice 2: Scalable streaming speech synthesis with large language models},
  author={Du, Zhihao and Wang, Yuxuan and Chen, Qian and Shi, Xian and Lv, Xiang and Zhao, Tianyu and Gao, Zhifu and Yang, Yexin and Gao, Changfeng and Wang, Hui and others},
  journal={arXiv preprint arXiv:2412.10117},
  year={2024}
}

@article{flowgrpo,
  title={Flow-grpo: Training flow matching models via online rl},
  author={Liu, Jie and Liu, Gongye and Liang, Jiajun and Li, Yangguang and Liu, Jiaheng and Wang, Xintao and Wan, Pengfei and Zhang, Di and Ouyang, Wanli},
  journal={arXiv preprint arXiv:2505.05470},
  year={2025}
}

@article{shao2024deepseekmath,
  title={Deepseekmath: Pushing the limits of mathematical reasoning in open language models},
  author={Shao, Zhihong and Wang, Peiyi and Zhu, Qihao and Xu, Runxin and Song, Junxiao and Bi, Xiao and Zhang, Haowei and Zhang, Mingchuan and Li, YK and Wu, Yang and others},
  journal={arXiv preprint arXiv:2402.03300},
  year={2024}
}

@inproceedings{reddy2022dnsmos835,
  title={DNSMOS P. 835: A non-intrusive perceptual objective speech quality metric to evaluate noise suppressors},
  author={Reddy, Chandan KA and Gopal, Vishak and Cutler, Ross},
  booktitle={Proc. ICASSP},
  pages={886--890},
  year={2022},
  organization={IEEE}
}

@inproceedings{speechbertscore,
  title     = {{SpeechBERTScore: Reference-Aware Automatic Evaluation of Speech Generation Leveraging NLP Evaluation Metrics}},
  author    = {Takaaki Saeki and Soumi Maiti and Shinnosuke Takamichi and Shinji Watanabe and Hiroshi Saruwatari},
  year      = {2024},
  booktitle = {Proc. Interspeech},
  pages     = {4943--4947},
  doi       = {10.21437/Interspeech.2024-1508},
  issn      = {2958-1796},
}

@article{li2025mixgrpo,
  title={Mixgrpo: Unlocking flow-based grpo efficiency with mixed ode-sde},
  author={Li, Junzhe and Cui, Yutao and Huang, Tao and Ma, Yinping and Fan, Chun and Yang, Miles and Zhong, Zhao},
  journal={arXiv preprint arXiv:2507.21802},
  year={2025}
}

@inproceedings{ERes2Net,
  title     = {{An Enhanced Res2Net with Local and Global Feature Fusion for Speaker Verification}},
  author    = {Yafeng Chen and Siqi Zheng and Hui Wang and Luyao Cheng and Qian Chen and Jiajun Qi},
  year      = {2023},
  booktitle = {Proc. Interspeech},
  pages     = {2228--2232},
  doi       = {10.21437/Interspeech.2023-1294},
  issn      = {2958-1796},
}

@inproceedings{dns2020,
  title     = {{The INTERSPEECH 2020 Deep Noise Suppression Challenge: Datasets, Subjective Testing Framework, and Challenge Results}},
  author    = {Chandan K.A. Reddy and Vishak Gopal and Ross Cutler and Ebrahim Beyrami and Roger Cheng and Harishchandra Dubey and Sergiy Matusevych and Robert Aichner and Ashkan Aazami and Sebastian Braun and Puneet Rana and Sriram Srinivasan and Johannes Gehrke},
  year      = {2020},
  booktitle = {Proc. Interspeech},
  pages     = {2492--2496},
  doi       = {10.21437/Interspeech.2020-3038},
  issn      = {2958-1796},
}

@inproceedings{libritts,
  title     = {LibriTTS: A Corpus Derived from LibriSpeech for Text-to-Speech},
  author    = {Heiga Zen and Viet Dang and Rob Clark and Yu Zhang and Ron J. Weiss and Ye Jia and Zhifeng Chen and Yonghui Wu},
  year      = {2019},
  booktitle = {Proc. Interspeech},
  pages     = {1526--1530},
  doi       = {10.21437/Interspeech.2019-2441},
  issn      = {2958-1796},
}

@INPROCEEDINGS{voicebank,
  author={Veaux, Christophe and Yamagishi, Junichi and King, Simon},
  booktitle={Proc. O-COCOSDA/CASLRE}, 
  title={The voice bank corpus: Design, collection and data analysis of a large regional accent speech database}, 
  year={2013},
  volume={},
  number={},
  pages={1-4},
  doi={10.1109/ICSDA.2013.6709856}}

@inproceedings{deepcluster,
  title={{Deep clustering: Discriminative embeddings for segmentation and separation}},
  author={Hershey, John R and Chen, Zhuo and Le Roux, Jonathan and Watanabe, Shinji},
  booktitle={Proc. ICASSP},
  pages={31--35},
  year={2016},
  organization={IEEE}
}

@inproceedings{ears,
  title     = {{EARS: An Anechoic Fullband Speech Dataset Benchmarked for Speech Enhancement and Dereverberation}},
  author    = {Julius Richter and Yi-Chiao Wu and Steven Krenn and Simon Welker and Bunlong Lay and Shinji Watanabe and Alexander Richard and Timo Gerkmann},
  year      = {2024},
  booktitle = {Proc. Interspeech},
  pages     = {4873--4877},
  doi       = {10.21437/Interspeech.2024-153},
  issn      = {2958-1796},
}

@inproceedings{demands,
  title={{The diverse environments multi-channel acoustic noise database (demand): A database of multichannel environmental noise recordings}},
  author={Thiemann, Joachim and Ito, Nobutaka and Vincent, Emmanuel},
  booktitle={Proceedings of Meetings on Acoustics},
  volume={19},
  number={1},
  year={2013},
  organization={AIP Publishing}
}

@inproceedings{wham,
  title     = {{WHAM!: Extending Speech Separation to Noisy Environments}},
  author    = {Gordon Wichern and Joe Antognini and Michael Flynn and Licheng Richard Zhu and Emmett McQuinn and Dwight Crow and Ethan Manilow and Jonathan Le Roux},
  year      = {2019},
  booktitle = {Proc. Interspeech},
  pages     = {1368--1372},
  doi       = {10.21437/Interspeech.2019-2821},
  issn      = {2958-1796},
}

@inproceedings{realman,
  author       = {Bing Yang and
                  Changsheng Quan and
                  Yabo Wang and
                  Pengyu Wang and
                  Yujie Yang and
                  Ying Fang and
                  Nian Shao and
                  Hui Bu and
                  Xin Xu and
                  Xiaofei Li},
  editor       = {Amir Globersons and
                  Lester Mackey and
                  Danielle Belgrave and
                  Angela Fan and
                  Ulrich Paquet and
                  Jakub M. Tomczak and
                  Cheng Zhang},
  title        = {RealMAN: {A} Real-Recorded and Annotated Microphone Array Dataset
                  for Dynamic Speech Enhancement and Localization},
  booktitle    = {Proc. NeurIPS},
  year         = {2024},
  url          = {http://papers.nips.cc/paper\_files/paper/2024/hash/bf8f6f5b017dc60d0c4e28a7a9a4ee7b-Abstract-Datasets\_and\_Benchmarks\_Track.html},
  bibsource    = {dblp computer science bibliography, https://dblp.org}
}

@ARTICLE{fsd50k,
  author={Fonseca, Eduardo and Favory, Xavier and Pons, Jordi and Font, Frederic and Serra, Xavier},
  journal={IEEE/ACM Transactions on Audio, Speech, and Language Processing}, 
  title={FSD50K: An Open Dataset of Human-Labeled Sound Events}, 
  year={2022},
  volume={30},
  number={},
  pages={829-852},
  doi={10.1109/TASLP.2021.3133208}}

@inproceedings{urgent2025,
  title     = {{Interspeech 2025 URGENT Speech Enhancement Challenge}},
  author    = {Kohei Saijo and Wangyou Zhang and Samuele Cornell and Robin Scheibler and Chenda Li and Zhaoheng Ni and Anurag Kumar and Marvin Sach and Yihui Fu and Wei Wang and Tim Fingscheidt and Shinji Watanabe},
  year      = {2025},
  booktitle = {Proc. Interspeech},
  pages     = {858--862},
  doi       = {10.21437/Interspeech.2025-1363},
  issn      = {2958-1796},
}

@INPROCEEDINGS{openslr2628,
  author={Ko, Tom and Peddinti, Vijayaditya and Povey, Daniel and Seltzer, Michael L. and Khudanpur, Sanjeev},
  booktitle={Proc. ICASSP}, 
  title={A study on data augmentation of reverberant speech for robust speech recognition}, 
  year={2017},
  volume={},
  number={},
  pages={5220-5224},
  doi={10.1109/ICASSP.2017.7953152}}

@inproceedings{fma,
  author       = {Micha{\"{e}}l Defferrard and
                  Kirell Benzi and
                  Pierre Vandergheynst and
                  Xavier Bresson},
  editor       = {Sally Jo Cunningham and
                  Zhiyao Duan and
                  Xiao Hu and
                  Douglas Turnbull},
  title        = {{FMA:} {A} Dataset for Music Analysis},
  booktitle    = {Proc. ISMIR},
  pages        = {316--323},
  year         = {2017},
  url          = {https://ismir2017.smcnus.org/wp-content/uploads/2017/10/75\_Paper.pdf},
  bibsource    = {dblp computer science bibliography, https://dblp.org}
}

@inproceedings{dit,
  title={Scalable diffusion models with transformers},
  author={Peebles, William and Xie, Saining},
  booktitle={Proc. ICCV},
  pages={4195--4205},
  year={2023}
}

@inproceedings{mpsenet,
  title     = {{MP-SENet: A Speech Enhancement Model with Parallel Denoising of Magnitude and Phase Spectra}},
  author    = {Ye-Xin Lu and Yang Ai and Zhen-Hua Ling},
  year      = {2023},
  booktitle = {Proc. Interspeech},
  pages     = {3834--3838},
  doi       = {10.21437/Interspeech.2023-1441},
  issn      = {2958-1796},
}

@INPROCEEDINGS{zipenhancer,
  author={Wang, Haoxu and Tian, Biao},
  booktitle={Proc. ICASSP}, 
  title={{ZipEnhancer: Dual-Path Down-Up Sampling-based Zipformer for Monaural Speech Enhancement}}, 
  year={2025},
  volume={},
  number={},
  pages={1-5},
  doi={10.1109/ICASSP49660.2025.10888703}}

@inproceedings{hifigan,
  author       = {Jungil Kong and
                  Jaehyeon Kim and
                  Jaekyoung Bae},
  title        = {{HiFi-GAN: Generative Adversarial Networks for Efficient and High Fidelity
                  Speech Synthesis}},
  booktitle    = {Proc. NeurIPS},
  year         = {2020},
  url          = {https://proceedings.neurips.cc/paper/2020/hash/c5d736809766d46260d816d8dbc9eb44-Abstract.html},
  bibsource    = {dblp computer science bibliography, https://dblp.org}
}
\end{spacing}

\end{document}